\documentclass[twocolumn]{aastex631}

\shorttitle{The 2022 reactivation of \srclong}
\shortauthors{Ibrahim et al.}

\def\xmm {\emph{XMM--Newton}}
\def\cxo {\emph{Chandra}}
\def\nustar {\emph{NuSTAR}}

\def\swift {\emph{Swift}}
\def\nicer {\emph{NICER}}
\def\int {\emph{INTEGRAL}}

\def\srclong{SGR\,J1935+2154}
\def\src{SGR\,J1935}

\def\torun{Toru\'n}

\def\flux {\mbox{erg cm$^{-2}$ s$^{-1}$}}
\def\lum {\mbox{erg s$^{-1}$}}
\def\nh {$N_{\rm H}$}

\def\ss {\mbox{s\,s$^{-1}$}}
\def\chisq {$\chi ^{2}$}
\def\rchisq {$\chi_{\nu} ^{2}$}

\usepackage{xspace}
\usepackage{hyperref}
\usepackage[nolist,nohyperlinks]{acronym}

\newcommand{\dmunit}{pc~cm$^{-3}$\xspace}
\newcommand{\dspsr}{{\tt DSPSR}\xspace}
\newcommand{\digifil}{{\tt digifil}\xspace}
\newcommand{\filterbank}{{\tt filterbank}\xspace}
\newcommand{\psradd}{{\tt psradd}\xspace}
\newcommand{\psrplot}{{\tt psrplot}\xspace}

\begin{document}

\title{An X-ray and radio view of the 2022 reactivation of the magnetar SGR\,J1935$+$2154}

\correspondingauthor{A. Y. Ibrahim}
\email{ibrahim@ice.csic.es}

\author[0000-0002-5663-1712]{A.~Y.~Ibrahim}
\affiliation{Institute of Space Sciences (ICE, CSIC), Campus UAB, Carrer de Can Magrans s/n, E-08193, Barcelona, Spain}
\affiliation{Institut d’Estudis Espacials de Catalunya (IEEC), Carrer Gran Capità 2-4, E-08034 Barcelona, Spain}

\author[0000-0001-8785-5922]{A.~Borghese}
\affiliation{Instituto de Astrofísica de Canarias, E-38205 La Laguna, Tenerife, Spain}
\affiliation{Departamento de Astrofísica, Universidad de La Laguna, E-38206 La Laguna, Tenerife, Spain}

\author[0000-0001-7611-1581]{F.~Coti~Zelati}
\affiliation{Institute of Space Sciences (ICE, CSIC), Campus UAB, Carrer de Can Magrans s/n, E-08193, Barcelona, Spain}
\affiliation{Institut d’Estudis Espacials de Catalunya (IEEC), Carrer Gran Capità 2-4, E-08034 Barcelona, Spain}

\author[0000-0002-0430-6504]{E.~Parent}
\affiliation{Institute of Space Sciences (ICE, CSIC), Campus UAB, Carrer de Can Magrans s/n, E-08193, Barcelona, Spain}
\affiliation{Institut d’Estudis Espacials de Catalunya (IEEC), Carrer Gran Capità 2-4, E-08034 Barcelona, Spain}

\author[0000-0001-5674-4664]{A.~Marino}
\affiliation{Institute of Space Sciences (ICE, CSIC), Campus UAB, Carrer de Can Magrans s/n, E-08193, Barcelona, Spain}
\affiliation{Institut d’Estudis Espacials de Catalunya (IEEC), Carrer Gran Capità 2-4, E-08034 Barcelona, Spain}

\author[0000-0001-9381-8466]{O.~S.~Ould-Boukattine}
\affiliation{ASTRON, Netherlands Institute for Radio Astronomy, Oude Hoogeveensedijk 4, 7991 PD Dwingeloo, The Netherlands}
\affiliation{Anton Pannekoek Institute for Astronomy, University of Amsterdam, Science Park 904, 1098 XH, Amsterdam, The Netherlands}

\author[0000-0003-2177-6388]{N. Rea}
\affiliation{Institute of Space Sciences (ICE, CSIC), Campus UAB, Carrer de Can Magrans s/n, E-08193, Barcelona, Spain}
\affiliation{Institut d’Estudis Espacials de Catalunya (IEEC), Carrer Gran Capità 2-4, E-08034 Barcelona, Spain}

\author[0000-0001-5116-6789]{S. Ascenzi}
\affiliation{Gran Sasso Science Institute, Viale F. Crispi 7,I-67100,L’Aquila
(AQ), Italy}
\affiliation{INFN – Laboratori Nazionali del Gran Sasso, I-67100, L’Aquila
(AQ), Italy}
\affiliation{INAF – Osservatorio Astronomico di Brera, via E. Bianchi 46,
23807, Merate (LC), Italy}

\author[0009-0001-3911-9266]{D. P. Pacholski} \affiliation{INAF—Istituto di Astrofisica Spaziale e Fisica Cosmica di Milano, via A. Corti 12, I-20133 Milano, Italy}
\affiliation{Dipartimento di Fisica G. Occhialini, Università degli Studi di Milano Bicocca, Piazza della Scienza 3, I-20126 Milano, Italy}

\author[0000-0003-3259-7801]{S.~Mereghetti} 
\affiliation{INAF—Istituto di Astrofisica Spaziale e Fisica Cosmica di Milano, via A. Corti 12, I-20133 Milano, Italy}

\author[0000-0001-5480-6438]{G. L. Israel} 
\affiliation{INAF—Osservatorio Astronomico di Roma, via Frascati 33, I-00078 Monteporzio Catone, Italy}

\author[0000-0002-6038-1090]{A. Tiengo}
\affiliation{Scuola Universitaria Superiore IUSS Pavia, Palazzo del Broletto, piazza della Vittoria 15, I-27100 Pavia, Italy}
\affiliation{INAF—Istituto di Astrofisica Spaziale e Fisica Cosmica di Milano, via A. Corti 12, I-20133 Milano, Italy}

\author[0000-0001-5902-3731]{A. Possenti}
\affiliation{INAF–Osservatorio Astronomico di Cagliari, Via della Scienza 5, 09047 Selargius, CA, Italy}

\author[0000-0002-8265-4344]{M. Burgay} 
\affiliation{INAF–Osservatorio Astronomico di Cagliari, Via della Scienza 5, I-09047 Selargius, Italy}

\author[0000-0003-3977-8760]{R. Turolla} 
\affiliation{Dipartimento di Fisica e Astronomia “Galileo Galilei”, Università di Padova, via F. Marzolo 8, I-35131 Padova, Italy}
\affiliation{Mullard Space Science Laboratory, University College London, Holmbury St. Mary, Dorking, Surrey RH5 6NT, UK}

\author[0000-0001-5326-880X]{S. Zane} 
\affiliation{Mullard Space Science Laboratory, University College London, Holmbury St. Mary, Dorking, Surrey RH5 6NT, UK}

\author[0000-0003-4849-5092]{P. Esposito} 
\affiliation{Scuola Universitaria Superiore IUSS Pavia, Palazzo del Broletto, piazza della Vittoria 15, I-27100 Pavia, Italy}
\affiliation{INAF—Istituto di Astrofisica Spaziale e Fisica Cosmica di Milano, via A. Corti 12, I-20133 Milano, Italy}

\author[0000-0001-9494-0981]{D.~G\"otz} 
\affiliation{AIM-CEA/DRF/Irfu/Département d’Astrophysique, CNRS, Université Paris-Saclay, Université de Paris Cité, Orme des Merisiers, F-91191 Gif-sur-Yvette, France}

\author[0000-0001-6278-1576]{S.~Campana} 
\affiliation{INAF--Osservatorio Astronomico di Brera, Via Bianchi 46, Merate (LC), I-23807, Italy}

\author[0000-0001-6664-8668]{F.~Kirsten}
\affiliation{Department of Space, Earth and Environment, Chalmers University of Technology, Onsala Space Observatory, 439 92, Onsala, Sweden}

\author[0000-0003-4056-4903]{M.\,P.~Gawro\'nski}
\affiliation{Institute of Astronomy, Faculty of Physics, Astronomy and Informatics, Nicolaus Copernicus University, Grudziadzka 5, 87-100 Toru\'n, Poland}

\author[0000-0003-2317-1446]{J.\,W.\,T.~Hessels}
\affiliation{ASTRON, Netherlands Institute for Radio Astronomy, Oude Hoogeveensedijk 4, 7991 PD Dwingeloo, The Netherlands}
\affiliation{Anton Pannekoek Institute for Astronomy, University of Amsterdam, Science Park 904, 1098 XH, Amsterdam, The Netherlands}

\begin{abstract}

Recently, the Galactic magnetar \srclong\ has garnered attention due to its emission of an extremely luminous radio burst, reminiscent of Fast Radio Bursts (FRBs).
\srclong\ is one of the most active magnetars, displaying flaring events nearly every year, including outbursts as well as short and intermediate bursts. Here, we present our results on the properties of the persistent and bursting X-ray emission from \srclong\, during the initial weeks following its outburst on October 10, 2022. The source was observed with \xmm\ and \nustar\ (quasi-)simultaneously during two epochs, separated by $\sim$5 days. 
The persistent emission spectrum is well described by an absorbed blackbody plus power-law model up to an energy of $\sim$25\,keV. No significant changes were observed in the blackbody temperature ($kT_{\rm BB}\sim$ 0.4 keV) and emitting radius ($R_{\rm BB}\sim$ 1.9 km) between the two epochs. However, we observed a slight variation in the power-law parameters. 
Moreover, we detected X-ray pulsations in all the datasets and derived a spin period derivative of $\dot{P} = 5.52(5) \times 10^{-11}$\,\ss. This is 3.8 times larger than the value measured after the first recorded outburst in 2014.
Additionally, we performed quasi-simultaneous radio observations using three 25--32-m class radio telescopes for a total of 92.5\,hr to search for FRB-like radio bursts and pulsed emission. However, our analysis did not reveal any radio bursts or periodic emission.

\end{abstract}


\begin{acronym}
    \acrodef{PF}[PF]{pulsed fraction}
\end{acronym}

\section{Introduction} 
\label{sec:intro}

Magnetars are a sub-group of isolated neutron stars with ultra-high magnetic fields of $B \approx 10^{14}-10^{15}$\,G, whose decay and instability are believed to be the main energy source of their emission \citep{duncan92}. Magnetars have spin periods $P$ that range between 0.3--12\,s and large spin down rates between $\dot{P} \sim 10^{-13}-10^{-11}$\,\ss, although magnetar-like emission has also been detected from peculiar pulsars that may not necessarily have $P$ and $\dot{P}$ falling within the aforementioned range \citep[e.g.,][]{rea10, rea16, archibald16}. Magnetars are persistent X-ray sources with luminosities of $L_{X} \approx 10^{31} - 10^{36}$\,\lum\ \citep[for reviews see e.g.,][]{turolla15, kaspi17, esposito21}. In addition, they are characterised by transient activities, which may affect the spectral and timing properties of the persistent emission. Based on their duration, these activities can be divided into short- and long-lived events. The former include bursts of tens/hundreds of milliseconds duration and giant flares lasting up to a few minutes, and reaching peak luminosities as high as 10$^{47}$\,\lum. The latter, known as outbursts, are sudden increases of the persistent X-ray flux by a factor of 10-1000, followed by a gradual decay over a period of months to years \citep[see e.g., the Magnetar Outburst Online Catalog\footnote{\url{http://magnetars.ice.csic.es/}},][]{cotizelati18}.

On 2014 July 5, the Burst Alert Telescope (BAT) on board the {\it Neil Gehrels Swift Observatory} \citep[\swift;][]{gehrels04} detected a short burst, leading to the discovery of a new magnetar, \srclong\ \citep[\src\ in the following;][]{stamatikos14}. Follow-up observations enabled the measurement of the source spin period $P \sim 3.24$\,s and spin-down rate of $\dot{P} \sim 1.43\times 10^{-11}$\,\ss. These values resulted in a surface dipolar magnetic field $B \sim 2.2\times10^{14}$\,G at the equator, confirming the magnetar nature of the source \citep{israel16}. 
The distance to the magnetar has been the focus of various works. Some of these studies associate \src\ with the supernova remnant G57.2$+$0.8, for which distances of 6.6$\pm$0.7\,kpc \citep{zhou20} and $\leq 10$\,kpc \citep{kozlova16} have been derived. On the other hand, other studies reported a distance of 4.4$^{+2.8}_{-1.3}$\,kpc, based on the analysis of an expanding dust-scattering ring associated with a bright X-ray burst \citep{mereghetti20}.\\
Since its discovery, \src\, has been a very active source, experiencing multiple outbursts in 2015, 2016 (twice) and 2020 \citep[see e.g.,][]{younes17, borghese20}, as well as frequent bursting episodes \citep[e.g.,][]{lin20}. Additionally, one day after the 2020 reactivation, a short and very bright, double-peaked radio burst (known as FRB\,200428) temporally coincident with a hard X-ray burst was observed \citep[][]{chime20, bochenek20, mereghetti20, ridnaia21, tavani21, li21}. This was the first time \src\ was detected in the radio band. The radio burst showed properties similar to those of Fast Radio Bursts (FRBs), providing strong evidence that magnetars may power at least a subgroup of FRBs.

On 2022 October 10--11, multiple short X-ray bursts were detected from \src\ by \int\,, \swift/BAT and other X-ray satellites indicating a reactivation of the source \citep[e.g.,][]{mereghetti22-gcn, palmer22-atel, ibrahim22-ATel}. Following this bursting activity, \nicer\ began observing the source and measured a persistent X-ray flux that was about one order of magnitude higher than the quiescent level \citep{younes22atel}. A new outburst had begun. Similarly to the 2020 outburst, radio bursts with X-ray counterparts were also observed during the initial stage of this outburst \citep[e.g.,][]{maan22-ATel, chime22-ATel, younes22-ATel}, but none as bright as FRB\,200428. 

Here, we report on the X-ray persistent and bursting emission properties of \src\ during the first weeks of the most recent active period, as well as on our searches for single pulses and pulsed emission in quasi-simultaneous radio observations. We first summarise the X-ray data analysis procedure in Section\,\ref{sec:Xobs}. We then present the timing
and spectral analysis, as well as a search for short bursts in Section\,\ref{sec:analysis}. In Section\,\ref{sec:radioobs}, we describe our radio observations. Finally, Section\,\ref{sec:discuss} presents a discussion of our findings.

\section{X-ray observations and data reduction}
\label{sec:Xobs}
We report on nearly simultaneous \xmm\, and \nustar\, observations, carried out between 2022 October 15 and 22. Data reduction was carried out using \textsc{heasoft} package \citep[v6.31;][]{heasoft14} and the Science Analysis Software (\textsc{SAS}\footnote{\url{https://www.cosmos.esa.int/web/xmm-newton/sas}}, v.19.1.0 \citealt{gabriel04}) with the latest calibration files.

Throughout this work, we adopted the coordinates reported by \citet{israel16}, i.e. R.A. = 19$^\mathrm{h}$34$^\mathrm{m}$55$\fs$598, decl. = +21$^{\circ}$53$^{\prime}$47$\farcs$79 (J2000.0), and the JPL planetary ephemeris DE\,200 to convert the photon arrival times to the Solar system barycenter. Additionally, to be consistent with our previous works \citep[e.g.,][]{bci+22}, we adopted a distance of 6.6\,kpc  \citep{zhou20} and quote all uncertainties at a 1$\sigma$ confidence level (c.l.).

\subsection{\xmm}
\xmm\, observed \src\, twice with the European Photon Imaging Camera (EPIC), for an exposure time of $\sim$\,40\,ks and $\sim$\,50\,ks for the first (ID:0902334101, between 2022 October 15, 19:48:48 UTC, and October 16, 12:06:17 UTC) and the second (ID:0882184001, 2022 October 22 between 03:22:56 and 22:12:09 UTC) observation, respectively. For each observation, the EPIC-pn \citep{struder01} was set in Small Window mode (time resolution of 5.7\,ms) while the EPIC-MOS1 and EPIC-MOS2 \citep{turner01} were set in Full Window mode (time resolution of 2.6\,s) and Timing mode (time resolution of 1.75\,ms), respectively. Following standard procedures, we filtered the event files for periods of high background activity, resulting in a net exposure of 39\,ks and 41\,ks for the first and the second pointings. No pile-up was detected. The source counts were extracted from a circle of radius 30 arcsec centered on the source and the background level was estimated from a 60-arcsec-radius circle far from the source, on the same CCD. In this study, our primary focus was on data collected with the EPIC-pn, because of its higher counting statistics owing to its larger effective area compared to that of the two MOS. However, we verified that the MOS data yielded consistent results.

\subsection{\nustar}
\src\, was observed twice with \nustar\, \citep{harrison13}: the first time between 2022 October 18, 21:51:09 UTC, and October 20, 22:21:09 UTC (ID:80702311002, on-source exposure time $\sim$50\,ks); the second time between 2022 October 22, 22:21:09 UTC, and October 24, 03:11:09 UTC (ID:80702311004, on-source exposure time $\sim$51\,ks). Source photons were accumulated within a circular region of radius 100 arcsec. A similar region centered on a position uncontaminated by the source emission was used for the extraction of the background events. The light curves, the spectra and the corresponding response files for the two focal plane detectors, referred to as FPMA and FPMB, were extracted using the \textsc{nuproducts} script.

\subsection{\int}
We searched the \int\ archive for data obtained simultaneously with \xmm\ and \nustar\ observations. This resulted in 23 pointings where \src\ was in the field of view of the IBIS coded mask imaging instrument. These pointings cover about $60\%$ of the first \xmm\ observation (from October 15 at 18:51 to October 16 at 04:47 UTC) and $15\%$ of the first \nustar\ observation (on October 19, from 14:43 to 17:45 UTC). We used data from the IBIS/ISGRI detector that operates in the nominal energy range 15--1000\,keV providing photon-by-photon data with excellent time resolution of 73\,$\mu$s. \int\ data were only examined for the presence of short bursts.

\section{X-ray Analysis and results}
\label{sec:analysis}

\subsection{X-ray timing analysis}
\label{sec:timing}

\begin{figure}
    \centering
    \includegraphics[scale=0.5]{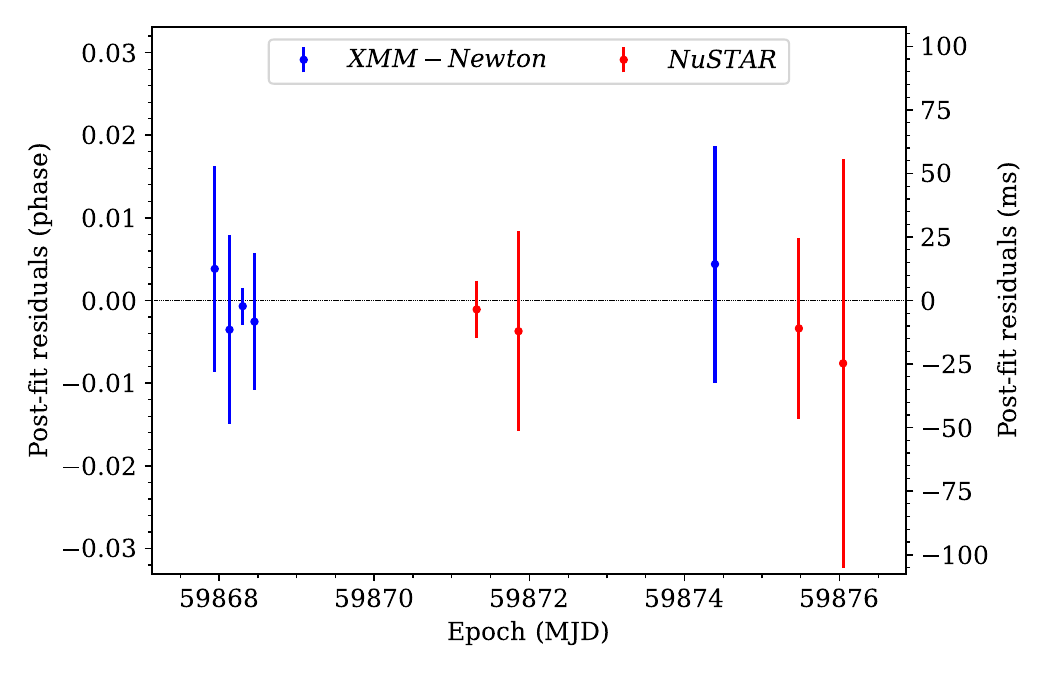}
    \caption{Post-fit residuals of our best-fit coherent timing solution for \src\ (Table~\ref{tab:timing}).  }
    \label{fig:timing-resids}
\end{figure}

To perform the timing analysis of \src, we first filtered out the burst events from the dataset so that they do not affect the integrated pulse profile morphology. We then used the {\tt{photonphase}} task of the {\tt{PINT}} software \citep{lsd+21} to assign a rotational phase to the barycentered events by extrapolating the ephemeris from \cite{bci+22}. In order to use the same fiducial reference phase for the \xmm\ and \nustar\ dataset, thus enabling phase coherence across the observations, only photons with energies below 15\,keV were analysed. We then combined those events into a stable template profile which we modeled with multiple Gaussian components. Using the {\tt{photon\_toa.py}} tool of the {\tt{NICERsoft}} package\footnote{\url{https://github.com/paulray/NICERsoft/wiki}}, we extracted barycentric pulse time of arrivals (TOAs) and proceeded to phase-connect the four dataset with the {\tt{TEMPO}} timing software \citep{nds+15}. We achieved coherence across the dataset using a simple model that only has the spin frequency $\nu$ and its first derivative $\dot{\nu}$ as free parameter. We show the post-fit residuals in Figure \ref{fig:timing-resids} and provide our coherent solution in Table~\ref{tab:timing}.

\begin{figure}
    \centering
    \includegraphics[scale=0.6]{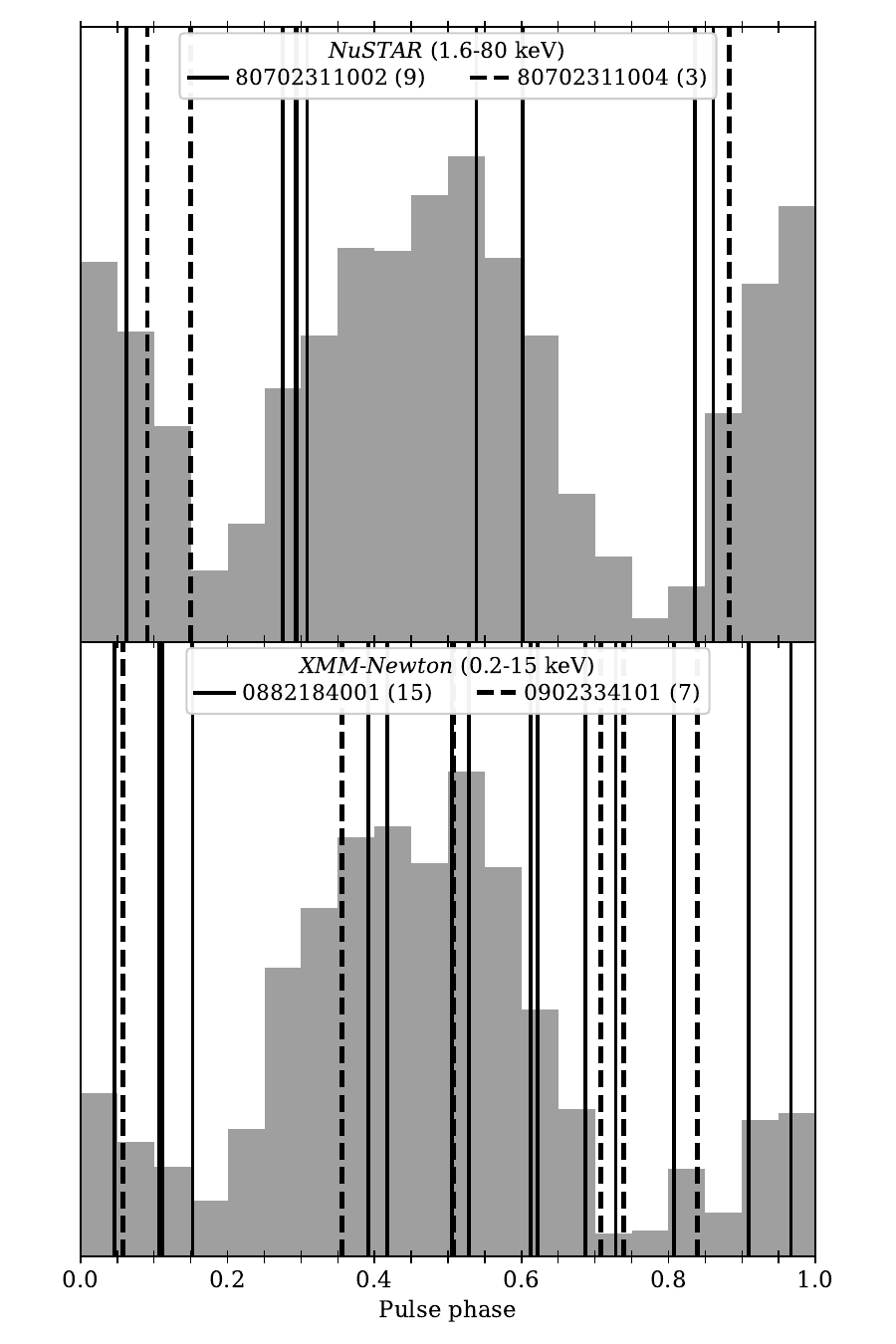}
    \caption{Phase distribution of the bursts (vertical black lines) detected in the \nustar\ (top) and \xmm\ (bottom) dataset (Table~\ref{tab:log_Xbursts}), plotted against the combined pulse profiles in each datasets (light grey) over one rotation cycle. The number of bursts in each observation is specified in parentheses next to the observation ID in the legends. The timing model of Table~\ref{tab:timing} was used for the absolute phase alignment. To show the burst phases more clearly, the burst widths (which have duty cycles ranging from $\sim$1 to 16\%) are not depicted in this figure. }
    \label{fig:burst-phase}
\end{figure}

Using our timing model, we then computed the rotational phase associated with the (barycentric) \xmm\ and \nustar\ burst epochs (Table~\ref{tab:log_Xbursts}). Figure~\ref{fig:burst-phase} shows the burst phases against the integrated pulse profiles observed with both instruments.  
We find no evidence for a preferred burst rotational phase: the burst cumulative distribution in phase across a full rotation cycle is statistically consistent with a uniform distribution (we determined a p-value $>$ 25\% using both an Anderson-Darling and Kolmogorov–Smirnov test). Similarly, \cite{younes20} found no obvious clustering at any particular phase for the $\sim$220 bursts emitted from \src\ during the 2020 reactivation.  

\begin{deluxetable}{lc}[ht]
  \tablewidth{0.4\columnwidth}
  \tabletypesize{\scriptsize}
  \label{tab:xray_timing_table}
  \tablecaption{Coherent timing solution of \src\ derived from the \xmm\ and \nustar\ data. Values in parentheses are the 1-$\sigma$ uncertainty in the last digit of the fitting parameters reported by \texttt{TEMPO}. The epoch of frequency refers to the reference time for the spin measurements at the Solar system barycenter, while the reference epoch is the phase-zero reference for TOA phase predictions. \label{tab:timing} }
  \tablecolumns{2}
  \tablehead{\colhead{Parameter}& \colhead{Measured Value}}
  \startdata
R.A. (J2000)                       &  19:34:55.598      \\
Decl. (J2000)                      &  21:53:47.79       \\
$\nu$ (s$^{-1}$)                   &  0.307525543(4)    \\
$\dot{\nu}$ ($10^{-12}$\,s$^{-2}$) &  -5.22(5)          \\
$P$ (s)                                 &  3.25176241(5)    \\
$\dot{P}$ ($10^{-11}$)       &  5.52(5)    \\
Epoch of frequency (MJD)           &  59871.00          \\
Validity range (MJD)               & 59867.9 -- 59876.0 \\
Reference epoch (MJD)              & 59871.320339421679 \\ [0.5em]
Timescale                          &  TDB                  \\
Solar system ephemeris             &  DE200                \\
RMS residuals (ms)                 &   10.8                \\ 
Daily-averaged RMS residuals (ms)  &    8.1                \\ [0.5em]
\hline \\ [-0.99em]
\phn & Derived Value \\ [0.25em]
\hline
Surface dipolar magnetic field, $B_{\rm eq}$ ($ 10^{14}$\,G) & 4.3 \\
Spin down luminosity, $\dot{E}$ ($ 10^{34}$\,erg\,s$^{-1}$) & 6.3 \\
Characteristic age, $\tau_c$ (yr) & 930 \\
  \enddata
\end{deluxetable}

\begin{figure*}
    \centering
    \includegraphics[scale=0.5,trim=2cm 3.2cm 2cm 4cm, clip=true]{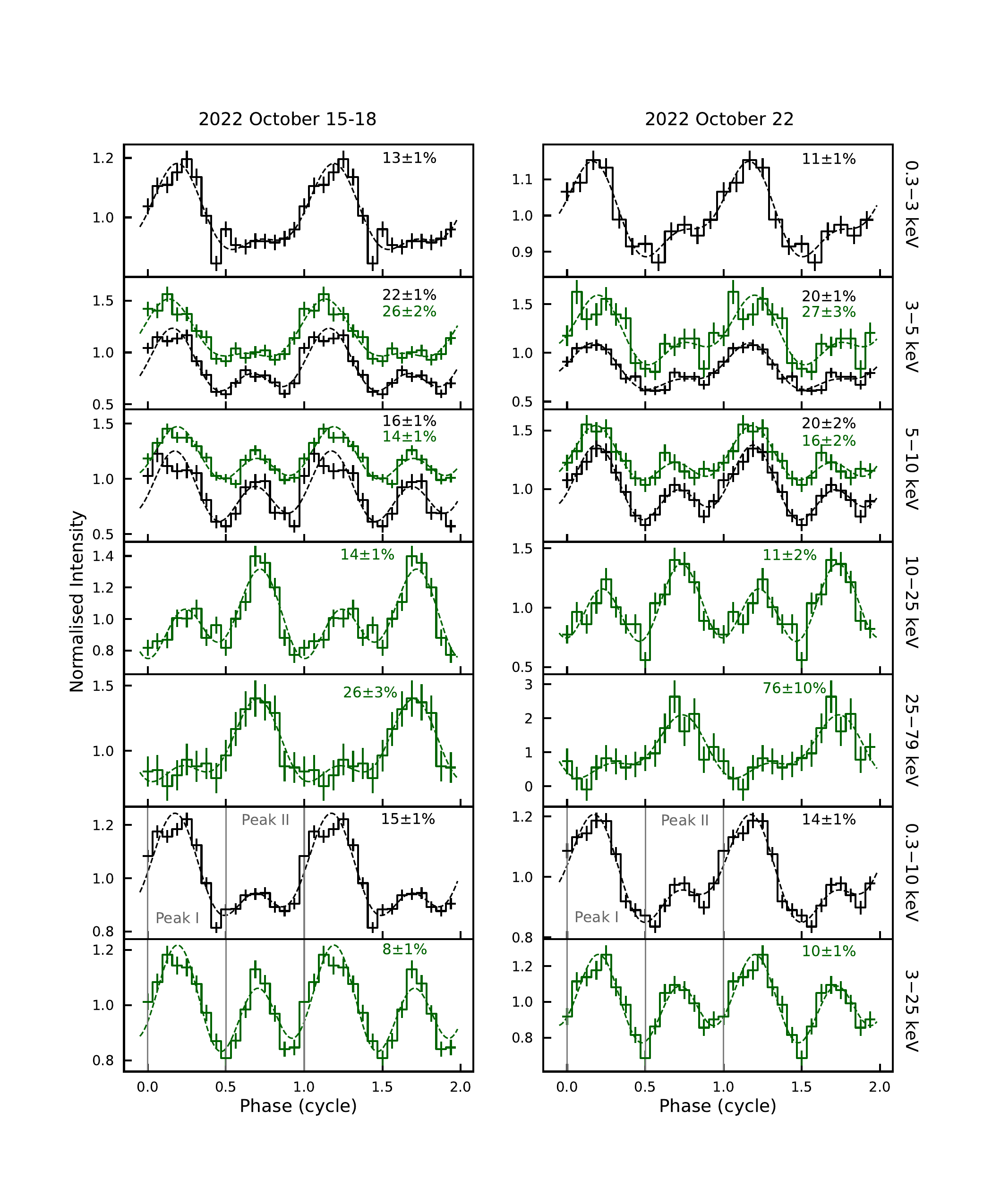}
    \caption{Background-subtracted, energy-resolved \xmm/EPIC-pn (black) and \nustar/FPMA+FPMB (green) pulse profiles for the 2022 October 15--18 (left-hand panel) and October 22 (right-hand panel) datasets. The dashed line in each panel indicates the best fit for the profiles (for more details, see Sec.~\ref{sec:timing}). The vertical grey lines in the last two panels denote the phase intervals adopted for the phase-resolved spectroscopy (for more details, see Sec.\ref{subsec:pps}). The corresponding pulsed fraction values are reported in each panel. Two cycles are shown for clarity and some pulse profiles have been arbitrarily shifted along the y-axis.}
    \label{fig:pulseprofile}
\end{figure*}

Figure~\ref{fig:pulseprofile} shows the background-subtracted light curves folded using the timing solution presented in Table~\ref{tab:timing} as a function of energy for the two epochs. We modelled all the pulse profiles with a combination of a constant plus two sinusoidal functions, with periods fixed to those of the fundamental and first harmonic components. The pulse profile exhibits a simple morphology below 3\,keV that evolves to a double-peaked shape at higher energies. At both epochs, the second peak (at phase $\sim$0.7) becomes more prominent above 10\,keV and dominates in the 25--79\,keV energy interval. The separation between the two peaks increases with energy for both epochs from $\sim$0.3--0.35 in phase at soft X-rays ($<$10\,keV) to $\sim$0.65--0.7 in phase at hard X-rays ($>$10\,keV). Moreover, we detected a phase shift $\Delta \phi$ between the soft (0.3--10\,keV) and hard (10--25\,keV) energy bands. For the first peak, $\Delta \phi_{0.3-10/10-25}$ is $0.13\pm0.02$ cycles during the first epoch, with the hard photons anticipating the soft ones, and it is not significant for the second epoch. While, for the second peak, we determined a shift of $\Delta \phi_{0.3-10/10-25}=0.19\pm0.01$ and $0.22\pm0.01$ cycles for the first and second epoch, respectively, with the soft photons leading the hard ones. Finally, we studied the dependence of the \ac{PF} with the photon energy and its time evolution. The \ac{PF} was computed by dividing the value of the semi-amplitude of the fundamental sinusoidal component describing the pulse profile by the average count rate. We did not detect any specific trend in the \ac{PF}, apart from (i) an increase between the 10--25\,keV and 25--79\,keV bands for both epoch, and (ii) an increase of the 25--79\,keV \ac{PF} between the two epochs.

\subsection{X-ray spectral analysis of the persistent emission and search for diffuse emission}
\label{sec:persistent}

The light curves of our observations exhibited several bursts, which will be properly investigated in Sec.\,\ref{sec:burst}. In order to exclude the bursts, we filtered out all the events with a count-rate higher than the average count-rate during the persistent state. We then used these filtered events to extract the spectra corresponding to the persistent emission only.

The spectral analysis was performed with \textsc{Xspec} \citep[v12.12.0;][]{arnaud96}. We used \textsc{specgroup} and \textsc{grppha} tools to group the spectra with a minimum of 50 counts per energy bin for \xmm/EPIC-pn and \nustar/FPMA datasets so as to use the \chisq\ statistics. 
In the following fits, we only used \nustar/FPMA spectra, but checked that \nustar/FPMB gave consistent results.
The \xmm\ spectra were fit in the 0.5--10\,keV energy interval, while for the \nustar\ ones the analysis was limited to the 3--25\,keV energy band owing to the low signal-to-noise ratio above 25\,keV. We adopted the \textsc{tbabs} model with chemical abundances from \citet{wilms00} and photoionization cross-sections from \citet{verner96} to describe the interstellar absorption.

\begin{figure*}
    \centering
    \includegraphics[width=1.03\columnwidth, trim= 0.35cm 0cm 0.25cm 0cm, clip=true]{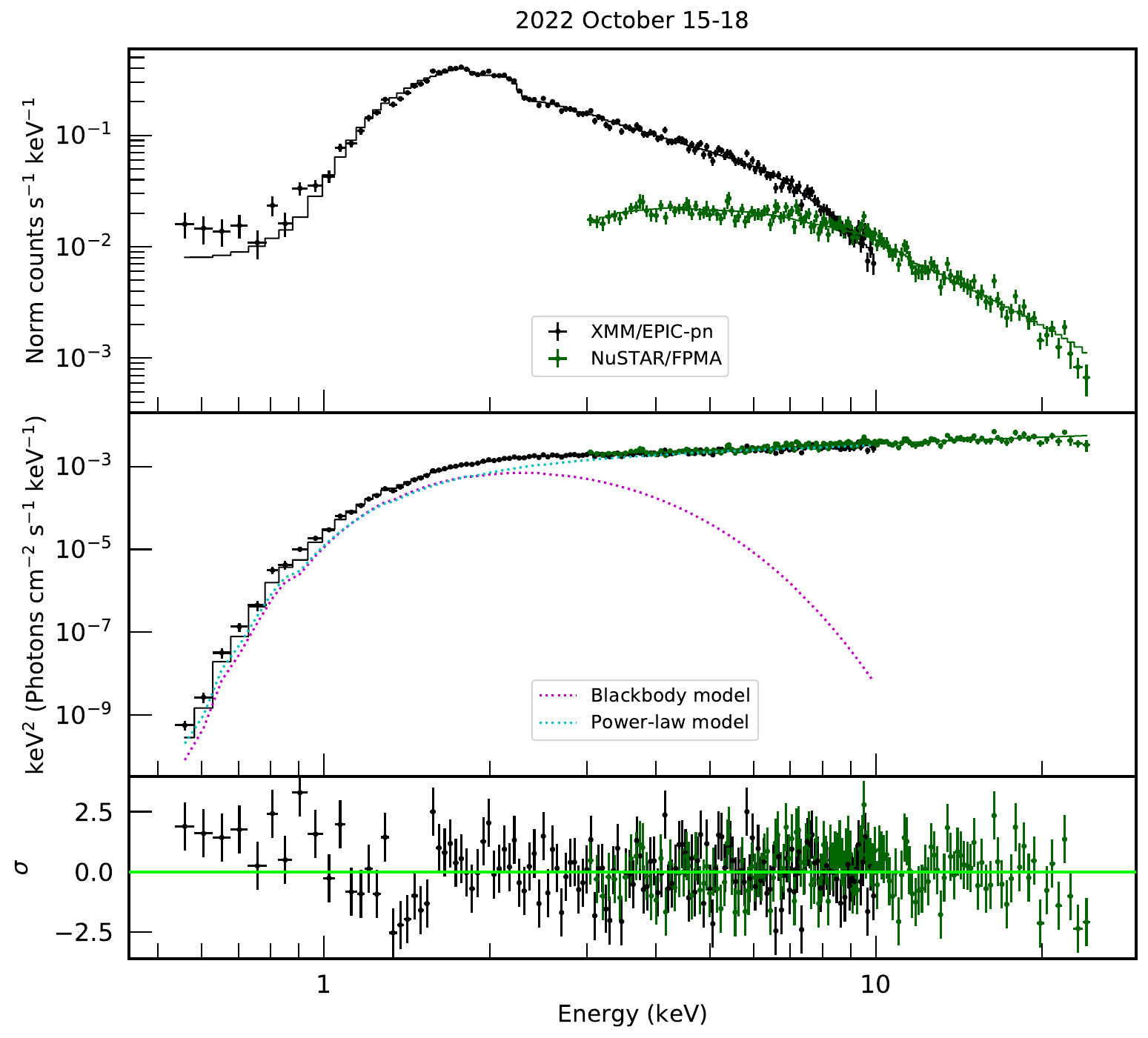}
    \includegraphics[width=1.\columnwidth, trim= 1.09cm 0cm 0.25cm 0cm, clip=true]{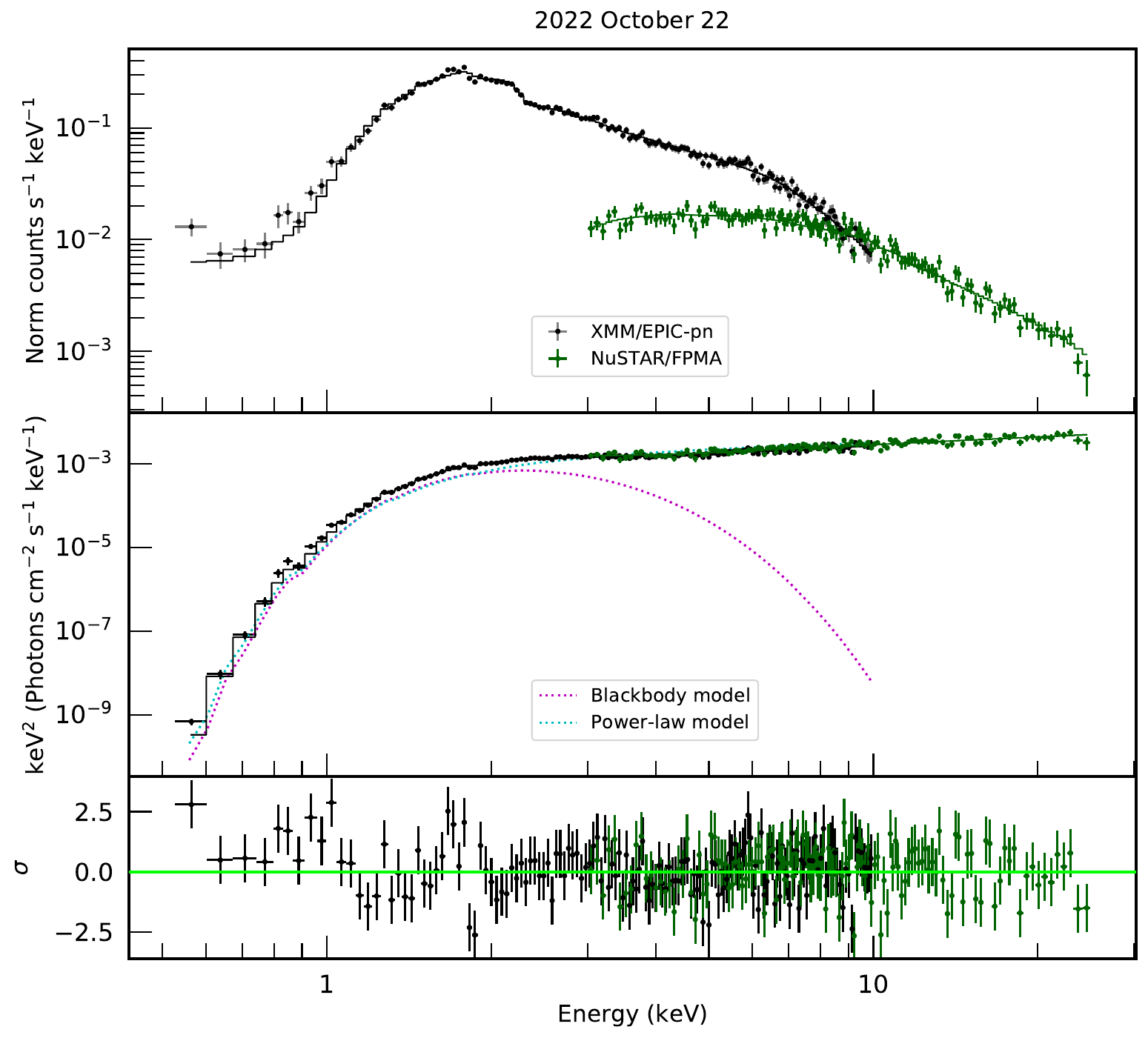} 
    \caption{Spectra of the persistent emission of \src. The 0.5--10\,keV \xmm/EPIC-pn (black) and the 3--25\,keV \nustar/FPMA (green) spectra are jointly fit with an absorbed blackbody plus power-law model. For each plot: the \textit{top panel} shows the counts spectra and the best-fitting model; the \textit{middle panel} shows the $E^2 f(E)$ unfolded spectra and the contribution of the single components (dotted lines); the \textit{bottom panel} shows the post-fit residuals in units of standard deviations.}
    \label{fig:persistent_spectra}
\end{figure*}

We simultaneously fit the \xmm\, and \nustar\, spectra with an absorbed blackbody plus power-law model (BB+PL), including a constant to account for cross-calibration between the two instruments  (see Figure\,\ref{fig:persistent_spectra}). \nh\ was tied up across all the four spectra, resulting in \nh $= (2.57 \pm 0.05)\times10^{22}$\,cm$^{-2}$ (reduced chi-square \rchisq=1.08 for 567 degrees of freedom (dof)). This value is compatible with those derived in previous studies of \src\, \citep[see e.g.,][]{younes17}.
For each epoch (2022 Oct 15–18 and 22), we linked all the BB+PL parameters across the \xmm\, and \nustar\ spectra. However, we allow these parameters to vary between the two epochs. 
Our analysis showed that there were no significant variations for the blackbody parameters between the first and second epoch, with an  emitting radius of $R_{\rm BB}\sim$1.9\,km and temperature of $kT_{\rm BB}\sim$0.4\,keV. On the other hand, the photon index slightly changed from $\Gamma=1.51\pm0.02$ to $1.41\pm0.02$ and the PL normalisation decreased by a factor of $\sim1.5$. The 0.5--25\,keV observed fluxes were $(1.26 \pm 0.02 )\times10^{-11}$ and $(1.04 \pm 0.02)\times10^{-11}$\,\flux, giving luminosities of $(9.17 \pm 0.07) \times10^{34}$ and $(7.48 \pm 0.07) \times10^{34}$\,\lum. The PL component accounted for $\sim93\%$ and $\sim89\%$ of the total luminosity at the first and second epochs, respectively. 

We also inspected the data taken from the EPIC-MOS1 detector for diffuse emission. For both epochs, we extracted radial profiles of the X-ray emission up to a distance of 100--150 arcsec from the magnetar, both from the images covering the entire observation duration, and from the images covering variable time intervals following the detection of the brightest X-ray bursts (see Sec.\,\ref{sec:burst} for more details). This second type of analysis was aimed at detecting short episodes of diffuse emission possibly associated with scattering haloes produced by the bursts. In no case did we find evidence of emission in excess of that from the magnetar.

\subsection{Phase-resolved spectroscopy}
\label{subsec:pps}

We performed a phase-resolved spectroscopy of the \xmm\ and \nustar\ datasets of the magnetar persistent emission. 
Our aim is to investigate any changes with rotational phase (and time) of the parameters of the spectra corresponding to the two pulse profile peaks. Therefore,
we extracted the 0.5--10\,keV EPIC-pn and 3--25\,keV FPMA spectra from the 0.0--0.5 (peak I) and 0.5--1.0 (peak II) phase intervals (see Figure \,\ref{fig:pulseprofile}).

The phase-resolved spectra were fit simultaneously with the BB+PL model. The column density was held fixed at the
phase-averaged value (\nh=2.57$\times10^{22}$\,cm$^{-2}$; see Sec.\,\ref{sec:persistent}). 
The spectral fitting results, reported in Table\,\ref{tab:spec_pps}, revealed variations along the spin phase, which can be primarily attributed to fluctuations in the PL photon index. During the first epoch, the variability was more pronounced with the index decreasing from 1.58$\pm$0.04 for peak I to 1.36$\pm$0.04 for peak II. In contrast, the second epoch displayed less variability with the index slightly changing from 1.30 $\pm$ 0.04 (peak I) to 1.43 $\pm$ 0.04 (peak II). At a given epoch, the BB parameters are consistent with each other in the different phase ranges.

\begin{table*}
\begin{center}
\caption{Results of the phase-resolved spectral analysis presented in Section \ref{subsec:pps}.}
\label{tab:spec_pps}
\begin{tabular}{lcccccc}
\hline
\hline
 &  \multicolumn{6}{c}{2022 Oct 15--18} \\
\hline
& Phase     & $kT_{\rm BB}$ & $R_{\rm BB}$ & $\Gamma$ & Flux$^a$ Unabs BB & Flux$^a$ Unabs PL \\
&          &  (keV)  &   (km)  &      & \multicolumn{2}{c}{(10$^{-12}$\,\flux)}  \\
\hline
Peak I & 0.0--0.5 & 0.42$\pm$0.02 & 1.3$\pm$0.1 & 1.58$\pm$0.04 & 1.38$\pm$0.02 & 7.36$\pm$0.01  \\
Peak II & 0.5--1.0 & 0.44$\pm$0.01 & 1.26$\pm$0.08 & 1.36$\pm$0.04 & 1.61$\pm$0.02  & 7.19$\pm$0.01 \\ 
\hline
\hline
 & \multicolumn{6}{c}{2022 Oct 22} \\
 \hline 
& Phase     & $kT_{\rm BB}$ & $R_{\rm BB}$ & $\Gamma$ & Flux$^a$ Unabs BB & Flux$^a$ Unabs PL \\
&          &  (keV)  &   (km)  &      & \multicolumn{2}{c}{(10$^{-12}$\,\flux)} \\
\hline
Peak I & 0.0--0.5 & 0.41$\pm$0.01 & 1.86$\pm$0.09 & 1.30$\pm$0.04 & 2.52$\pm$0.01 & 12.79$\pm$0.01\\
Peak II & 0.5--1.0 & 0.41$\pm$0.01 & 1.83$\pm$0.09 & 1.43$\pm$0.04 & 2.38$\pm$0.01 &  10.05$\pm$0.01\\
\hline
\hline
\label{tab:spectral_analysis}
\end{tabular}
\begin{list}{}{}
\item[$^a$]The fluxes are estimated in the 0.5--25\,keV energy range.
\end{list}
\end{center}
\end{table*}

\subsection{X-ray burst search and properties}
\label{sec:burst}

We investigated the \xmm\ and \nustar\ light curves of all observations for the presence of short bursts, applying the method described by \citet{borghese20} (see also, e.g., \citealt{gkw04}). We extracted time series with three different time resolutions (1/16, 1/32 and 1/64\,s) in order to identify events of different durations. We classified a time bin as a burst if it had a probability $<$10$^{-4}$($NN_{\rm trials}$)$^{-1}$ of being a Poissonian fluctuation of the average count rate, where $N$ is the total number of time bins in a given light curve and $N_{{\rm trials}}$ is the number of timing resolutions used in the search. We detected a total of 22 and 12 bursts in the \xmm/EPIC-pn and merged \nustar/FPMA+FPMB light curves, respectively. The burst epochs referred to the Solar system barycenter, as well as the burst fluences and durations, are reported in Table\,\ref{tab:log_Xbursts} and Figure\,\ref{fig:burst_LC} shows the light curves for the two strongest bursts detected in \xmm\ and \nustar\ data.

We extracted the spectra for those events with at least 25 net counts for \xmm\ and for the event with the highest counting statistics for \nustar\ (i.e., the burst labelled 80702311002\,\#9 in Table\,\ref{tab:log_Xbursts} with 80 net counts). The background level was estimated from time intervals of the same duration in the persistent state. 
We employed a minimum number of counts to group the spectra that varies from burst to burst depending on the fluence of the burst itself. We applied the chi-square statistic for model fitting, except for the cases where the counting statistic was too low. In such cases, we adopted the $W$-statistic instead.
The spectra were fitted with an absorbed blackbody model, fixing \nh\ to the value obtained from the analysis of the phase-average broadband spectrum. The fit results are reported in Table\,\ref{tab:log_Xbursts}.

Furthermore, for each observation, we extracted a stacked spectrum of all bursts and assigned the spectrum of the persistent-only emission as the background spectrum. We then fit the stacked spectra using the same model we adopted for the spectra of the single bursts (i.e., an absorbed blackbody with \nh\ fixed at $2.57 \times10^{22}$\,cm$^{-2}$). The \xmm\ spectra were well described by a single blackbody with temperature of $kT_{\rm BB}=1.14\pm0.06$\,keV and $kT_{\rm BB}=1.88\pm0.08$\,keV for the first and second epochs, respectively. Using the assumed distance of \src, i.e 6.6\,kpc, we obtained radii of $R_{\rm BB}=0.9\pm0.1$\,km for the first epoch and $R_{\rm BB}=1.14\pm0.07$\,km for the second one. However, this model was unsatisfactory for the \nustar\ spectra, and thus a second blackbody component was added. This resulted in temperatures of $kT_{\rm BB,cold}=0.5\pm0.2$\,keV and $kT_{\rm BB,hot}=3.1\pm0.3$\,keV for the cold and hot components, respectively, with radii of $R_{\rm BB,cold} = 8^{+39}_{-3}$\,km and $ R_{\rm BB,hot} = 0.27^{+0.06}_{-0.04}$\,km for the first epoch. For the second epoch, the temperatures were $kT_{\rm BB,cold}=0.8\pm0.3$\,keV and $kT_{\rm BB,hot}=4^{+4}_{-1}$\,keV with radii of $R_{\rm BB,cold} = 1.7^{+6.6}_{-0.5}$\,km and $R_{\rm BB,hot} = 0.09\pm0.03$\,km.

For the \int\ data, the burst search was carried out in the 30--150 and 30--80\,keV energy ranges, by examining light curves binned on seven timescales between 10 and 640\,ms. Only the pixels that had more than 50\% of their surface illuminated by the source were considered in our analysis.
Potential bursts were identified as significant excesses above the expected background level derived from a running average. Once identified, these excesses were then examined through an imaging analysis to confirm their authenticity and positional association with the magnetar. This search resulted in the detection of only two bursts.

Among the three bursts seen with \xmm\ during the \int\ observations (i.e., the bursts labelled 0902334101\,\#1, \#2 and \#3 in Table\,\ref{tab:log_Xbursts}), only the brightest one (\#3) was detected by \int\ as well. The burst had a fluence of $36.6$ counts (30--150\,keV) in ISGRI, over a duration of about 90\,ms. The light curve is shown in Figure\,\ref{fig:burst_LC}. We assume a spectrum described by thermal bremsstrahlung with a temperature of 30\,keV, which is commonly used to describe spectrum of magnetar bursts \citep[e.g.][]{borghese19}. The resulting average count rate of $406.6$ counts\,s$^{-1}$ corresponds to a flux of $2.04\times10^{-8}$\,\flux. 
The two bursts detected by \nustar\ (8070231100\,\#7 and \#8) were not visible in the \int\ data. The second burst detected with ISGRI  occurred on 2022 October 19 at 15:25:54.037 (UTC), during a time gap in the \nustar\ data. Its fluence and duration were 49 counts (30--150\,keV) over 200\,ms. The rate of 245.0 counts\,s$^{-1}$ corresponds to a flux of $1.23\times10^{-8}$\,\flux. 
 
\section{Quasi-simultaneous radio observations}
\label{sec:radioobs}
We observed \src\ using three radio telescopes in Europe: the 25-m RT-1 telescope in Westerbork, the~Netherlands (Wb), the 25-m telescope in Onsala, Sweden (O8) and the 32-m telescope in \torun, Poland (Tr). Observations were carried out at $1.4$\,GHz, $1.6$\,GHz (L-band) and $330$\,MHz (P-band) (see Table \ref{tab:radio-dish} for the observational setup). The source was monitored between October 15 and 19, 2022 for a total of 92.5\,hr. This number reduces to 60.4\,hr when taking into account the overlap between observations at different telescopes.  

\begin{table*}
\centering
\caption{\label{tab:radio-dish}Observational setup of the radio telescopes.}
\begin{tabular}{cccccccc}
\hline
\hline
Station$\mathrm{^{a}}$  & Band & Frequency Range & Bandwidth$\mathrm{^{b}}$ & Bandwidth per & SEFD$\mathrm{^{c}}$ & Completeness$\mathrm{^{d}}$ & Time observed \\
& & [MHz] & [MHz] & subband [MHz] & [Jy] & [Jy~ms] & [hrs] \\
\hline
Wb  & P                 & 300--364          &50     & 8           & 2100  & 46     & 11.4\\
Wb  & L    & 1207--1335        &100    & 16          & 420   & 7     & 45.5 \\
Tr  & L    & 1350--1478         &100   & 16          & 250   & 4     & 22.0 \\
O8  & L$_{\rm O8-1}$    & 1360--1488        &100    & 16          & 310   & 5     & 6.3 \\
O8  & L$_{\rm O8-2}$    & 1594.49--1722.49        &100    & 16          & 310   & 5     & 7.4 \\
\hline
\multicolumn{7}{l}{Total telescope time/total time on source [hrs]$\mathrm{^{e}}$} & 92.5/60.4 \\
\hline
\multicolumn{8}{l}{$\mathrm{^{a}}$ Wb: Westerbork RT1 25-m, O8: Onsala 25-m, Tr: Toru\'n 32-m} \\
\multicolumn{8}{l}{$\mathrm{^{b}}$ Effective bandwidth accounting for RFI and band edges.} \\
\multicolumn{8}{l}{$\mathrm{^{c}}$ From the \href{http://old.evlbi.org/user_guide/EVNstatus.txt}{EVN status page}.} \\
\multicolumn{8}{l}{$\mathrm{^{d}}$ Using Equation \ref{eq:radiometer}, assuming a $7\sigma$ detection threshold and a pulse width of $1$~ms.} \\
\multicolumn{8}{l}{$\mathrm{^{e}}$ Total time on source accounts for overlap between the participating stations.} \\
\end{tabular}
\end{table*}

\subsection{Single pulse search}

We searched the data for FRB-like emission applying the custom pipeline described by \citet{Kirsten_2021_NatAs, kirsten_2022_natur}. 

Data is recorded as ``raw voltages", also known as baseband data, at each station in \textit{.vdif} format \citep{whitney_2010_ivs}. This format encapsulates dual circular polarization with 2-bit sampling. In order to search the data, we first create Stokes I (full intensity) \filterbank files with 8-bit encoding using \digifil which is part of \dspsr \citep{vanstraten_2011_pasa}. For observations at L-band, the frequency resolution is $125$\,KHz, and the time resolution of the \filterbank is $64~\mu s$, with the exception of Tr, which has a time resolution of $8~\mu s$. For the P-band observation, these values are $512~\mu s$ and $7.8125$\,KHz, respectively. We mitigated radio frequency interference (RFI) by applying a static mask. This mask is manually determined for each station and observational setup by identifying channels affected by RFI. We then searched the data for burst candidates using Heimdall\footnote{\url{https://sourceforge.net/projects/heimdall-astro/}}, setting a signal-to-noise threshold of $7$. We only searched for bursts within a dispersion measure (DM) range of $\pm~50$ units, with the known DM of \src\ being $332.7206 \pm 0.0009$~\dmunit \citep{chime20}. Burst candidates are subsequently classified using the machine learning classifier FETCH \citep{agarwal_2020_mnras}.
We use models A \& H and set a probability threshold of $50\%$. The produced burst candidates were then all manually inspected to determine if they are astrophysical or RFI.

\subsection{Search for pulsed emission}

In an effort to detect pulsed radio emission from \src, we folded our radio data using the ephemeris derived from the X-ray data (see Sec.\,\ref{sec:timing}). Additionally, we also folded individual scans which were coincident with an X-ray burst. Overall, we had six instances of overlap between X-ray burst detections and radio coverage. Four of these instances were covered by multiple radio telescopes simultaneously (see Table\,\ref{tab:emission} for details). 

The radio observations are divided into scans each lasting typically $900$\,s. We first identified the scan that encompassed an X-ray burst, as well as the scans immediately before and after it, totalling roughly $2700$\,s of data. We used \dspsr to fold the data based on the ephemeris. Folding was only possible due to the contemporaneous X-ray and radio observations. These folded scans were subsequently combined into a single file using \psradd. We then created a diagnostic plot using \psrplot to determine the presence of pulsed emission. We validated this method by applying it to observations of the pulsar J1935+1616.

\subsection{Results}
\label{sec:radio_results}
No FRB-like bursts were found in the radio observations. This allows us to calculate a completeness threshold. The completeness threshold is the upper limit on the fluence of a burst that falls below the sensitivity of our instruments and can be derived using the radiometer equation,

\begin{equation}\label{eq:radiometer}
    \mathcal{F} = (\textrm{S/N}) \cdot \frac{T_{\textrm{sys}}}{G} \cdot \sqrt{\frac{W}{n_{\textrm{pol}}\Delta \nu}} \ [\textrm{Jy ms}]~,
\end{equation}

\noindent where $(\textrm{S/N})$ is the signal-to-noise detection threshold value, $\frac{T_{\textrm{sys}}}{G}$ is the System-Equivalent Flux Density (SEFD), $W$ is the width of the burst, $n_{\textrm{pol}}$ is the number of recorded polarizations and $\Delta \nu$ is the recorded bandwidth. Using Equation\,\ref{eq:radiometer} and the properties of the radio telescopes listed in Table\,\ref{tab:radio-dish}, and assuming a width of 1~ms and a 7$\sigma$ detection threshold, we can find completeness thresholds of 5\,Jy\,ms for Onsala, 4\,Jy\,ms for \torun, 7\,Jy\,ms and 46\,Jy\,ms for Westerbork L- and P-band, respectively. 
Moreover, we folded radio data at the times of overlap between X-ray detections of bursts and we folded all recorded L-band data spread over four days from Westerbork and \torun, which corresponds to $45.5$\,hr and $21.9$\,hr of observations, respectfully. We found no evidence for pulsed radio emission from \src\ using both approaches. We can therefore determine an upper limit on the typical minimum flux density using the following equation:

\begin{equation}\label{eq:meanflux}
    S_{\textrm{mean}} = (\textrm{S/N}) \cdot \frac{\beta T_{\textrm{sys}}}{G\sqrt{n_{\textrm{pol}}t_{\textrm{obs}}\Delta \nu}} \cdot \sqrt{\frac{W}{P-W}} \ [\textrm{Jy}]~,
\end{equation}

\noindent where $\beta$ is a factor accounting for quantization effects and is approximated to be $1.1$ \citep[see][and references therein]{lorimer_2004_hpa}; $P$ is the spin period of the source as quoted in Table\,\ref{tab:xray_timing_table}; and $W$ is the width of the folded profile which is assumed to be equal to $10\%$ of the period. 
A complete overview of all derived upper limits can be found in Table\,\ref{tab:emission}. For the Westerbork P-band observation we find a mean flux density limit of $14.86$~mJy, while for the L-band observations we find flux density limits between $0.23-2.1$~mJy for the different telescopes, configurations and integration times. 

\section{Discussion} 
\label{sec:discuss}

On 2022 October 10-11, the magnetar \src\ entered a new outburst, characterized by the emission of several short X-ray bursts and an increase of the persistent X-ray flux. Moreover, like the previous outburst in 2020, the source emitted a few radio bursts with X-ray counterparts \citep[e.g.,][]{younes22-ATel}. This event is the sixth detected outburst from \src, making this magnetar one of the most active known so far. 

Here, we presented the properties of the X-ray persistent emission and bursts of \src\ during the first weeks of its most recent outburst based on observations obtained with \xmm\ and \nustar.
Additionally, we performed searches for single pulses and pulsed emission through quasi-simultaneous radio observations without any successful results.\\

\noindent {\bf Flux and spectral decomposition}: \\
The outburst onset was marked by the emission of several short X-ray bursts between 10 and 11 October 2022 \citep[see e.g.,][]{palmer22-atel,mereghetti22-gcn}. Our observations were carried out $\sim$6 and 12 days later. At both epochs, emission was detected up to 25\,keV (see Fig.\,\ref{fig:persistent_spectra}). Hard X-ray emission from \src\ was also seen in a pointing performed $\sim$5 days after the 2015 outburst onset and was still observed 5 months after the 2020 reactivation \citep{younes17,bci+22}. The persistent X-ray spectra were well modeled by the combination of a thermal and non-thermal components. The thermal component was well described by a blackbody model. Its parameters remained stable over time, with a temperature of $\sim$0.4\,keV and radius of $\sim$1.9\,km.
The non-thermal component had a power-law shape and its contribution to the total 0.5--25\,keV luminosity decreased only marginally from $\sim$93\% to $\sim$89\% in about 5 days. 

The quiescent level of \src\ is not known yet. Here, we adopt the quiescent observed flux derived by \cite{bci+22} using a \xmm\ observation performed on 2014 October 4, i.e. $(8.7\pm0.3)\times10^{-13}$\,\flux\ (0.3--10\,keV). The ratio between the 0.3--10\,keV observed flux measured during our first observation, $(6.45\pm0.05)\times10^{-12}$\,\flux, and that in quiescence is $R_{\rm 2022}\sim7.4$. Assuming the same quiescent flux and considering the peak fluxes of the previous outbursts measured by \cite{younes17} and \cite{borghese20}, we calculated the same ratio. Upon comparison, we found that $R_{\rm 2022}$ was greater than the values from the 2014 and 2015 events, which were $R_{\rm 2014} \sim4.9$ and $R_{\rm 2015} \sim5.4$, respectively. However, it was lower than the ratios from the May and June 2016 outbursts, which were $R_{\rm 2016May} \sim9.7$ and $R_{\rm 2016June} \sim16$, respectively.
Notably, the 2020 reactivation was the most powerful, with a ratio of $R_{\rm 2020} \sim49$.\\

\noindent {\bf Spin-down rate and pulse profile}: \\
We detected the spin period and the spin-down rate using \xmm\ and \nustar\ datasets, covering the period of 15--22 October 2022. We were able to establish a phase-coherent timing solution (see Table\,\ref{tab:timing}). 
The spin-down rate we inferred was markedly different from those derived during previous outbursts. Specifically, our results indicated that the spin-down rate during the first weeks on the 2022 reactivation ($\dot{P} \simeq 5.52(5) \times 10^{-11}$\,\ss) was a factor of 3.8 times larger than the value measured during the first four months of the 2014 outburst ($\dot{P} \simeq 1.43 \times 10^{-11}$\,\ss; \citealt{israel16}), and 1.5 times larger than the spin-down rate during the 2020 outburst ($\dot{P} \simeq 3.5 \times 10^{-11}$\,\ss; \citealt{bci+22}, see also \citealt{younes20}, \citealt{younes2023}). The observed variations in the spin-down rate suggest a notable change in the factors affecting the spin-down, e.g. the magnetospheric geometry and/or the relativistic wind of \src\ during different outbursts.
Moreover, changes in the spin-down rate are common during outbursts, indicating changes in the magnetosphere caused by the rearrangement of magnetic fields. To determine the secular spin-down rate of \src, a targeted monitoring campaign during the quiescence state is needed.
The evolution of the pulse profile during the 2022 reactivation of \src\ displays some differences when compared to previous outbursts. 
The pulse profiles observed in both \xmm\ and \nustar\ observations exhibits  a distinctive double-peaked morphology (see Fig.\,\ref{fig:pulseprofile}). Notably, the second peak (at phase $\sim$0.7) becomes more prominent at energies above 10\,keV for both epochs. The observed double-peaked structure contrasts with the quasi-sinusoidal shape showed during the 2014 outburst, as reported in \xmm\ and \cxo\ observations \citep{israel16}. 
However, it closely resembles that extracted from \nustar\ and \xmm\ observations taken during the 2020 outburst \citep{borghese20, bci+22}. The change of the pulse profile from a single-peak shape in the 2014 outburst to a double-peak shape during the 2022 reactivation may be related to the fact that different regions on the neutron star surface are heated during each outburst. Similarly to the 2014 outburst, we detected an energy-dependent pulse profile phase shift. Slight phase shifts between the peak emissions in the soft and hard X-ray pulse profiles have been observed in a number of magnetars (see e.g., XTE\,J1810$-$197 \citep{borghese21}, 1E\,1547.0$-$5408 \citep{cotizelati20}, and references therein). This phenomenology is consistent with the widely accepted scenario that magnetars non-thermal X-ray emission stems from resonant inverse Compton scattering of photons emitted from the star surface by charged particles moving along magnetic loops anchored to the crust and corotating with the star \citep[][and references therein]{wadiasingh18}. In this scenario, the hard, non-thermal X-ray emission is expected to be beamed along the loop and to be misaligned (in most cases) to some extent with respect to the soft, thermal X-ray emission pattern from the hot spots on the star surface. The \ac{PF} increased when shifting from the 10--25\,keV to 25--79\,keV energy bands at each epochs. We also observed a time-dependent change in the \ac{PF} for the 25--79\,keV and 3--25\,keV energy intervals with its value increasing between the two epochs. These results are inconsistent with the findings reported by \cite{israel16}, where they reported a time independent \ac{PF} in the 17--21\% range.\\

\noindent {\bf Pulse profile modelling}: \\
We determine the emission geometry of \src\ by examining the orientation of the hot spot relative to the line of sight and the star's rotational axis. To achieve this, we compared the observed \ac{PF} to a set of simulated \acp{PF} generated using the method outlined by \citet{Perna2001} and \citet{gotthelf10}.

Our approach involved creating a temperature map on the surface of the star. This map included a uniform background temperature and a single hot spot characterized by a Gaussian temperature profile. The hot spot's orientation with respect to the star's rotational axis was defined as an angle $\chi$, while we also specified the line of sight's orientation as an angle $\psi$ relative to the rotational axis. We then computed the observed phase-resolved spectra by integrating the local blackbody emission from the visible part of the stellar surface. In this calculation, we considered the effects of gravitational light bending, approximating the ray-tracing function \citep{Pechenick1983, Page1995} using the formula derived by \citet{beloborodov02}. Additionally, we took into account absorption by the interstellar medium. Since our model includes thermal emission only, we restrict our analysis to the energy range 0.3--2\,keV where the blackbody component dominates the emission. In this range, the \ac{PF} is $10.8\pm1.4\,\%$ in the first epoch, and $7.3\pm1.1\,\%$ in the second one. The pulse profile can be modelled using a simple sinusoidal function with a single peak per rotational phase, so in our modelling we consider a temperature map with a single hot-spot. For the temperature and the radius of the hot-spot, we considered the values obtained from the phase-resolved spectral-fit of peak I reported in Table\,\ref{tab:spectral_analysis}. The contribution from the rest of the stellar surface is neglected since it does not contribute significantly to the emission.

\begin{figure}
    \centering
\includegraphics[width = \columnwidth]{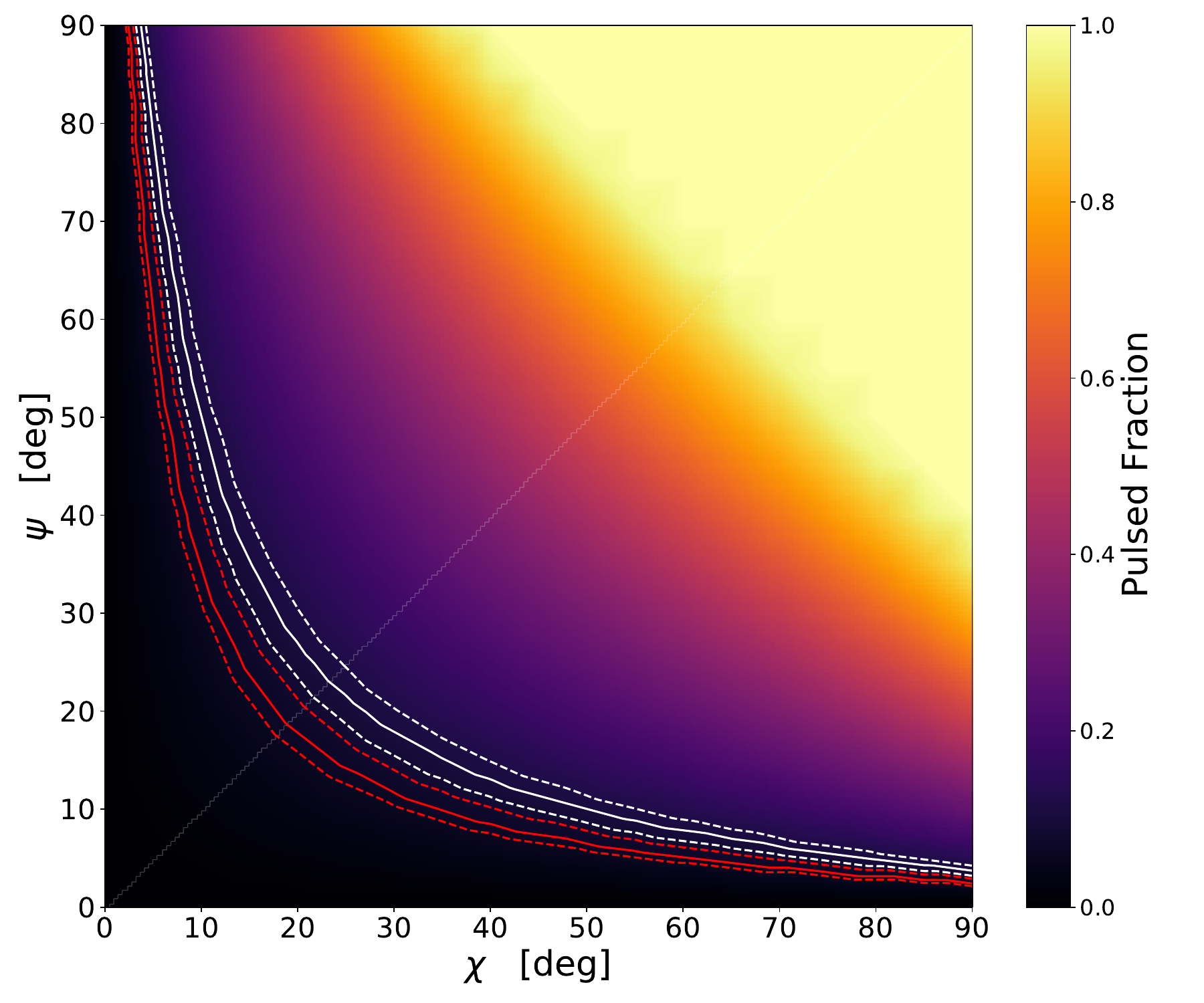}
    \caption{Constraints on the emission geometry of \src, based on the \ac{PF} measured in the first epooch (15th October 2022). The color scale represents the 0.3--2\,keV PF at different angles. The white lines represent the measured value ($\textrm{PF} = 10.8\pm1.4 \%$), while the red lines represent the measured value at the second epoch ($\textrm{PF} = 7.3\pm1.1 \%$). }
    \label{fig:PF_geometry}
\end{figure}

We report the results of our analysis in Figure\,\ref{fig:PF_geometry}. The color map on the $\chi-\psi$ plane represents the value of the \ac{PF} obtained by our modelling using the input parameters from the first epoch. The white and red contours represent the regions matching the observed \ac{PF} in the first and second epoch, respectively. Continuous curves represent the central value of the \ac{PF}, dashed curves represents the $1\sigma$ uncertainty regions. While the two regions do not overlap, they are consistent within 2$\sigma$.  
Our analysis suggests two preferable configurations: one where both angles have moderate values (\emph{e.g.} $(\chi-\psi)\sim (25^\circ - 25^\circ)$) and another where the line-of-sight is near the rotational axis and the hot-spot is almost perpendicular to it.

\begin{acknowledgments} 
AYI's work has been carried out within the framework of the doctoral program in Physics of the Universitat Autònoma de Barcelona. AYI, FCZ, EP, AM and NR are supported by the H2020 ERC Consolidator Grant “MAGNESIA” under grant agreement No. 817661 (PI: Rea) and Catalan grant SGR-Cat 2021 (PI: Graber). AB acknowledges support from the Consejer\'ia de Econom\'ia, Conocimiento y Empleo del Gobierno de Canarias and the European Regional Development Fund (ERDF) under grant with reference ProID2021010132 ACCISI/FEDER, UE. FCZ is supported by a Ramon y Cajal fellowship.
This work was also partially supported by the program Unidad de Excelencia Mar\'ia de Maeztu CEX2020-001058-M. Research by the AstroFlash group at University of Amsterdam, ASTRON and JIVE is supported in part by an NWO Vici grant (PI Hessels; VI.C.192.045). This work was supported by the NWO XS grant: WesterFlash (OCENW.XS22.1.053; PI: Kirsten). Part of this work has been funded using resources from the INAF Large Grant 2022 “GCjewels” (P.I. Andrea Possenti) approved with the Presidential Decree 30/2022. This work acknowledges support from Onsala Space Observatory for the  provisioning of its facilities/observational support. The Onsala Space Observatory national research infrastructure is funded through Swedish Research Council grant No 2017-00648. This work makes use of data from the Westerbork Synthesis Radio Telescope owned by ASTRON. ASTRON, the Netherlands Institute for Radio Astronomy, is an institute of the Dutch Scientific Research Council NWO (Nederlandse Oranisatie voor Wetenschappelijk Onderzoek). We thank the Westerbork operators R.Blauw, J.J. Sluman and H. Mulders for scheduling observations. This work is based in part on observations carried out using the 32-m radio telescope operated by the Institute of Astronomy of the Nicolaus Copernicus University in Toru\'n (Poland) and supported by a Polish Ministry of Science and Higher Education SpUB grant.

\end{acknowledgments}

\appendix
\restartappendixnumbering

\section{Log of short X-ray bursts}

Table~\ref{tab:log_Xbursts} lists the epochs, fluence, durations, best-fit spectral parameters and unabsorbed fluxes for the bursts detected in our datasets. 
The fluence refers to the 3--79\,keV and 0.2--12\,keV ranges for \nustar\ and \xmm\ bursts, respectively. The duration has to be considered as an approximate value. We estimated it by summing the 15.625-ms time bins showing enhanced emission for the structured bursts, and by setting it equal to the coarser time resolution at which the burst is detected in all the other cases.

\begin{deluxetable*}{rrcccccc}[htb]
\tabletypesize{\scriptsize}
\tablecaption{Log of X-ray bursts detected in all datasets and results of the spectral analysis for the brightest events. The \nh\ has been fixed to the average value in the spectral fits.
\label{tab:log_Xbursts}}
\tablecolumns{8}
\tablewidth{0pt}
\tablehead{
\colhead{Instrument/Obs.ID\tablenotemark{\footnotesize a}} &
\colhead{Burst epoch} &
\colhead{Fluence} &
\colhead{Duration}	&
\colhead{$kT_{\rm BB}$}	&
\colhead{$R_{\rm BB}$}	&
\colhead{$F_{\rm X,unabs}$\tablenotemark{\footnotesize b}} &
\colhead{\chisq / $W$-stat (dof)}\\
\colhead{} &
\colhead{YYYY-MM-DD hh:mm:ss (TDB)} &
\colhead{(counts)} & 
\colhead{(ms)} &
\colhead{(keV)} &
\colhead{(km)} &
\colhead{($\times10^{-9}$\,\flux)} &
}
\startdata
XMM/0902334101 \#1$^\dag$      & 2022-10-15 20:26:14.457 & 17 & 31.25  &			&	&	&  \\ 
\#2$^\dag$                     		& 2022-10-16 00:41:42.870 & 11 & 62.5 	&				&	&	&   \\
\#3$^\star$                     		&  03:53:09.083 & 55 & 109.375 		& 1.5$\pm$0.2			& $3.0_{-0.6}^{+0.8}$	& 0.9$\pm$0.1	& \chisq=15.86 (14)   \\
\#4                     		&  10:35:28.285 & 31 & 62.5 			& $1.7_{-0.4}^{+0.7}$  	& $7.6_{-2.4}^{+5.2}$	& 10$\pm$3 	& $W$-stat=21.55 (11)   \\
\#5                     		&  10:45:11.000 & 10 & 62.5 			&				&	&	&   \\
\#6                     		&  10:45:14.351 & 61 & 109.375		& $2.2_{-0.5}^{+0.8}$  	& $4.6_{-1.2}^{+2.1}$	& 9$\pm$2	& \chisq=5.14 (6)   \\ \vspace{0.2cm}
\#7                    			 &  12:05:02.934 & 29 & 62.5 			& $1.4_{-0.2}^{+0.4}$  	& $7.7_{-2.0}^{+3.7}$	& 5$\pm$1	& $W$-stat=13.54 (16)   \\ 
\nustar/80702311002 \#1 & 2022-10-19 06:29:29.769 & 25 & 46.875 &		&	&	&    \\
\#2                     		& 07:56:58.869 & 13 & 125 			&		&	&	&   \\
\#3                     		& 08:21:05.061 & 8 & 62.5 			&		&	&	&   \\
\#4                     		& 09:48:56.934 & 21 & 46.875 			&		&	&	&   \\
\#5$^\dag$                     		& 11:33:02.606 & 20 & 46.875 			&		&	&	&   \\
\#6$^\dag$                     		& 13:21:31.841 & 30 & 62.5 			&		&	&	&   \\
\#7$^\dag$                     		& 17:24:38.512 & 12 & 31.25 			&		&	&	&   \\
\#8$^\dag$                     		& 17:46:13.429 & 15 & 125 			&		&	&	&   \\ \vspace{0.2cm}
\#9                     		& 2022-10-20 00:13:17.634 & 80 & 171.875 &	3.1$^{+0.6}_{-0.4}$	& 1.0$^{+0.8}_{-0.6}$	& 1.2$\pm$0.2	&  $W$-stat=10.87 (17)  \\
XMM/0882184001 \#1     & 2022-10-22 03:59:47.011 & 16 & 62.5 	&		&	&	&   \\
\#2                     		& 04:27:31.542 & 9  & 31.25 			&		&	&	&   \\
\#3                     		& 04:46:13.754 & 110 & 218.75 		& $2.2_{-0.3}^{+0.4}$ 	& $3.9_{-0.7}^{+1.0}$	& 5.9$\pm$0.8  	& \chisq=4.15 (6)   \\
\#4                     		& 04:53:17.448 & 20 & 62.5 			&		&	&	&   \\
\#5                     		& 05:01:16.104 & 14 & 62.5 			&		&	&	&   \\
\#6                     		& 06:12:48.464 & 20 & 125 			&		&	&	&   \\
\#7                     		& 06:18:35.417 & 28 & 93.75 			& $2.6_{-0.7}^{+1.8}$	& $3.2_{-1.1}^{+2.3}$	& 7$\pm$2	& $W$-stat=14.42 (13)   \\
\#8                     		& 09:29:20.325 & 27 & 93.75 			& $1.9_{-0.4}^{+0.7}$	& $4.9_{-1.4}^{+2.9}$	& 6$\pm$2	& $W$-stat=11.01 (14)   \\
\#9                     		& 10:01:26.472 & 132 & 187.5 			& $2.3_{-0.4}^{+0.6}$	& $4.0_{-0.9}^{+1.4}$	& 7$\pm$1	& \chisq=7.33 (6)   \\
\#10                    		& 14:18.57.919 & 27 & 125 			& $1.4_{-0.3}^{+0.4}$ 	& $3.1_{-0.8}^{+1.8}$	& 0.8$\pm$0.2	& \chisq=2.74 (4)   \\
\#11                    		& 15:41:35.417 & 12 & 62.5 			&		&	&	&   \\     
\#12                    		& 16:25:01.920 & 30 & 156.25 			& $2.4_{-0.6}^{+1.2}$	& $2.9_{-0.9}^{+1.7}$	& 4$\pm$1	& $W$-stat=13.08 (18)   \\
\#13                    		& 16:31:33.816 & 123 & 203.125 		& $1.9_{-0.2}^{+0.3}$ 	& $4.8_{-0.9}^{+1.3}$	& 5.3$\pm$0.7	& \chisq=14.13 (8)  \\
\#14                    		& 16:42:44.030 & 28 & 125 			& $0.8_{-0.1}^{+0.2}$	& $12.9_{-3.7}^{+8.0}$	& 1.5$\pm$0.4	& $W$-stat=4.91 (8)   \\ \vspace{0.2cm}
\#15                    		& 17:37:26.814 & 290 & 531.25 		& 2.1$\pm$0.2			& $3.4_{-0.4}^{+0.5}$	& 4.0$\pm$0.3 	& \chisq=21.87 (24)    \\ 
\nustar/80702311004 \#1 & 2022-10-22 22:57:23.582 & 23 & 62.5 	&		&	&	&   \\
\#2                     		& 2022-10-23 21:58:05.838 & 10 & 62.5 	&		&	&	&   \\
\#3                     		& 22:50:23.135 & 27 & 62.5 			&		&	&	&   \\
\enddata
\tablenotetext{a}{The notation \#N corresponds to the burst number in a given observation.}
\tablenotetext{b}{The flux was estimated in the 0.5--10\,keV range for \xmm\ and \nustar.}
\tablenotetext{^\dag}{These bursts were covered by radio observations (for details, see Table\,\ref{tab:emission}).}
\tablenotetext{^\star}{Burst detected also with \int.}
\end{deluxetable*}


\begin{figure*}
    \centering
    \includegraphics[width=.6\columnwidth]{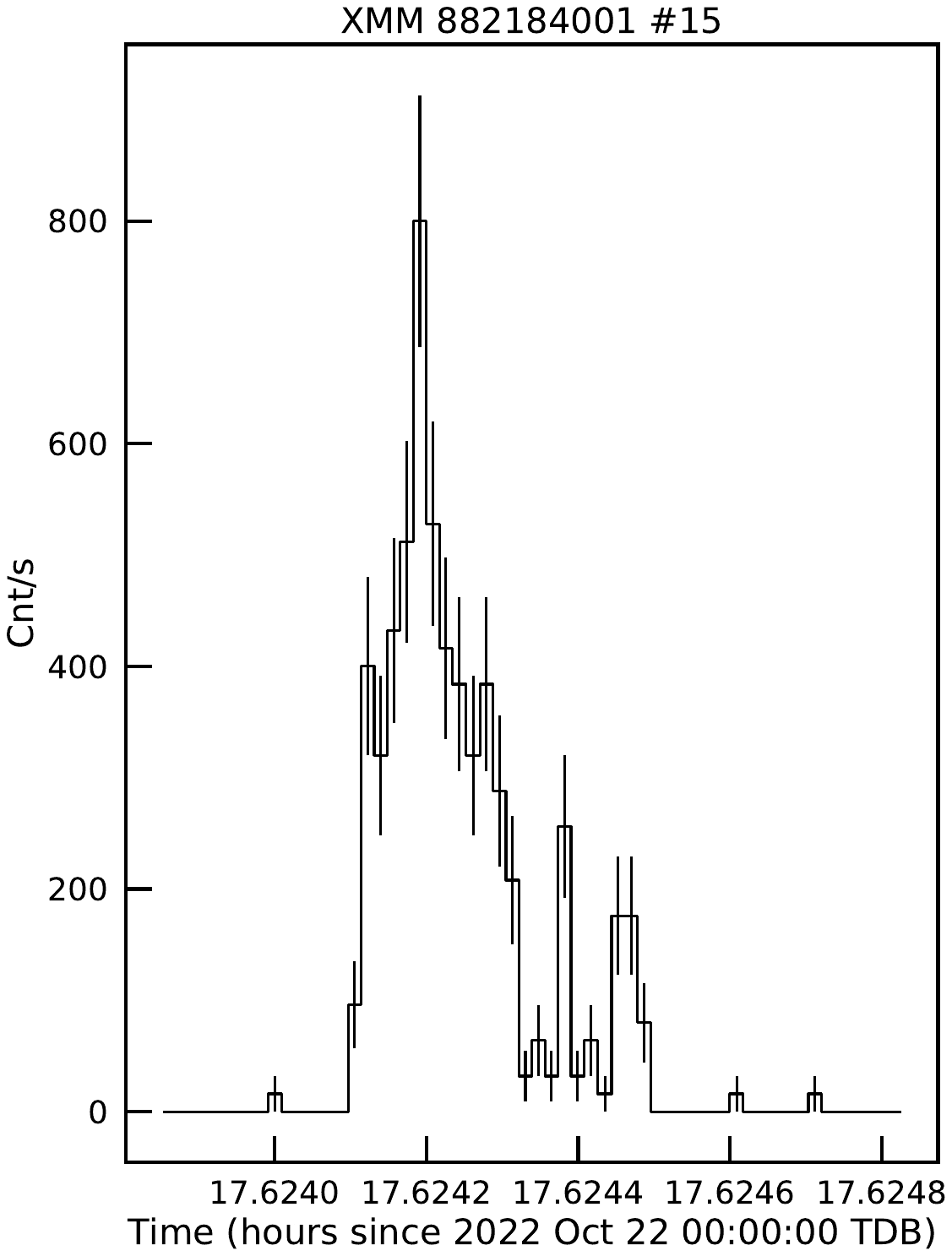}
    \includegraphics[width=.6\columnwidth]{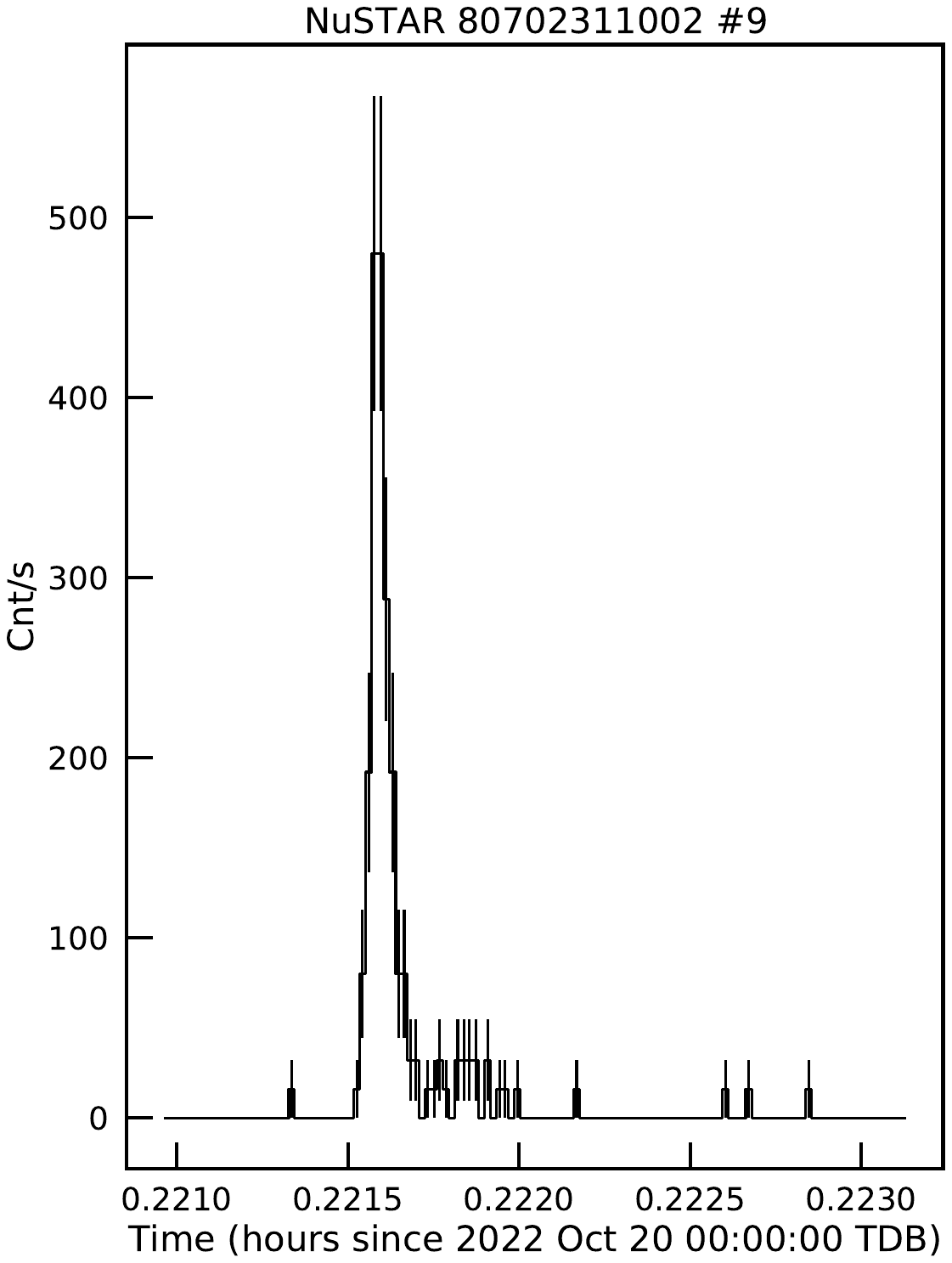}
    \includegraphics[width=.6\columnwidth]{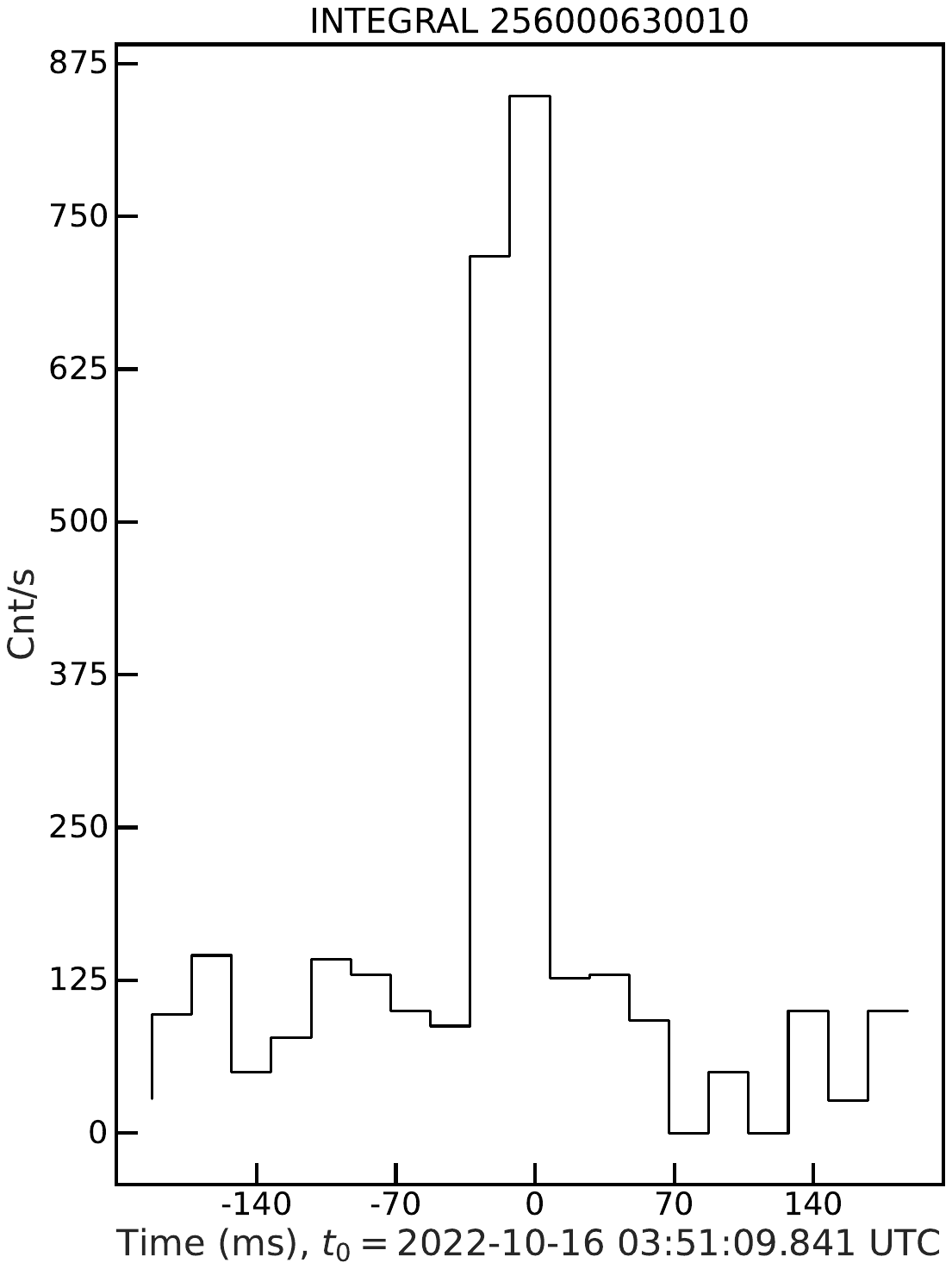}
     \vspace{-0.6cm}
    \caption{Light curves extracted from \xmm/EPIC-pn (left-hand panel) and \nustar/FPMA+FPMB (middle panel) data for the strongest bursts, binned at 62.5\,ms, while \int/IBIS/ISGRI (right-hand panel) data is binned at 20\,ms.}
    \label{fig:burst_LC}
\end{figure*}

\begin{table*}
\centering
\caption{\label{tab:emission} Limits on the mean flux density $S{\mathrm{_{mean}}}$ after folding the radio data for the entire Westerbork and \torun\ observations using the ephemeris as derived in the X-ray analysis. Additionally, we also fold and place upper limits on the flux density in the case of X-ray burst overlap instances.}
\begin{tabular}{rccccccc}
\hline
\hline
Overlap X-ray & Station & Band & Start time$\mathrm{^{a}}$ & Stop time$\mathrm{^{a}}$ & \#Scans & Exposure time & $S{\mathrm{_{mean}}}$$\mathrm{^{b}}$\\
& & & [TOPO UTC] & [TOPO UTC] & & [s] & [mJy] \\
\hline
                        & Tr    & L                 & 2022-10-15 14:30:08    & 2022-10-19 22:11:59   & 111   & 79041   & 0.23    \\
                        & Wb    & L                 & 2022-10-16 11:30:41    & 2022-10-19 23:14:38   & 180   & 163754   & 0.27   \\
\hline
XMM/0902334101 \#1      & Wb    & P                 & 2022-10-15 20:13:19    & 2022-10-15 20:58:38   & 3     & 2685  & 14.86     \\
\#1                     & O8    & L$_{\rm O8-2}$    & 2022-10-15 20:19:37    & 2022-10-15 21:04:58   & 3     & 2685   & 1.55     \\
\#1                     & Tr    & L                 & 2022-10-15 20:12:57    & 2022-10-15 20:49:49   & 3     & 2138   & 1.40     \\
\#2                     & O8    & L$_{\rm O8-1}$    & 2022-10-16 00:24:14    & 2022-10-16 00:54:21   & 2     & 1791   & 1.90     \\
NuSTAR/80702311002 \#5  & Wb    & L                 & 2022-10-19 11:18:42    & 2022-10-19 12:04:04   & 3     & 2690   & 2.10     \\
\#6                     & Wb    & L                 & 2022-10-19 13:00:50    & 2022-10-19 13:46:10   & 3     & 2687   & 2.10     \\
\#7                     & Wb    & L                 & 2022-10-19 17:04:03    & 2022-10-19 17:49:24   & 3     & 2690   & 2.10     \\
\#7                     & Tr    & L                 & 2022-10-19 17:12:37    & 2022-10-19 17:49:29   & 3     & 2137   & 1.40     \\
\#8                     & Wb    & L                 & 2022-10-19 17:34:27    & 2022-10-19 18:19:48   & 3     & 2691   & 2.10     \\
\#8                     & Tr    & L                 & 2022-10-19 17:37:38    & 2022-10-19 18:14:30   & 3     & 2136   & 1.40     \\
\hline
\multicolumn{8}{l}{$\mathrm{^{a}}$The time elapsed between start and stop times is not continuous due to $\sim10$-s gaps between scans.} \\
\multicolumn{8}{l}{$\mathrm{^{b}}$Using Equation \ref{eq:meanflux}, properties from Table \ref{tab:radio-dish} and assuming a $10\sigma$ detection and $10\%$ duty cycle.} \\
\end{tabular}
\end{table*}

\facilities{XMM-Newton, NuSTAR, INTEGRAL, Westerbork, Onsala, \torun}
\software{HEASoft (v6.31; \citealt{heasoft14}), FTOOLS (v6.27; \citealt{blackburn95}), XSPEC (v12.3.0; \citealt{arnaud96}),
NuSTARDAS (v1.9.2;\,\url{https://heasarc.gsfc.nasa.gov/docs/nustar/analysis/}), NICERsoft packag (\url{https://github.com/paulray/NICERsoft}), MATPLOTLIB (v3.6.2; \citealt{hunter07}), NUMPY (v1.23.5; \citealt{harris20}),
SAS (v20.0; \citealt{gabriel04}), TEMPO (\citealt{nds+15})}


\bibliography{biblio}{}
\bibliographystyle{aasjournal}

\end{document}